\documentclass[11pt,twoside]{article}

\usepackage{a4wide}
\usepackage{fullpage}
\usepackage{epsfig}
\usepackage{amsfonts}
\usepackage{amssymb}
\usepackage{amstext}
\usepackage{amsmath}
\usepackage{xspace}
\usepackage{shadow}
\usepackage{times}
\usepackage{theorem}
\usepackage{mathrsfs}
\usepackage{color}

\usepackage[compact]{titlesec}

\newtheorem{thm}{Theorem}
\newtheorem{cor}[thm]{Corollary}
\newtheorem{lem}[thm]{Lemma}
\newtheorem{cl}[thm]{Claim}

\newtheorem{defn}[thm]{Definition}

\def\proof{\noindent {\bf Proof:}\hspace{2mm}}
\def\bsq{$\blacksquare$}

\def\dr{{Dial-a-Ride}\xspace}
\def\dks{dense-$k$-subgraph\xspace}
\def\kfp{$k$-forest problem\xspace}
\def\kmp{$k$-MST problem\xspace}
\def\poly{\text{poly}\xspace}
\def\sse{\subseteq}
\newcommand\ignore[1]{}

\newcounter{note}[section]

\title{Dial a Ride from $k$-forest}
\author{
Anupam Gupta\thanks{Computer Science Department, Carnegie Mellon
    University. Supported in part by an NSF CAREER
  award CCF-0448095, and by an Alfred P.\ Sloan Fellowship.}
\and MohammadTaghi Hajiaghayi\thanks{Computer Science Department,
Carnegie Mellon
    University. Supported in part by NSF ITR
grant
  CCR-0122581 (The ALADDIN project).}
\and Viswanath Nagarajan\thanks{Tepper School of Business, Carnegie
Mellon University. Supported in part by NSF grants CCF-0430751 and
ITR grant CCR-0122581 (The ALADDIN project).}
  \and R. Ravi$^{\ddagger}$
   }

\begin{document}
\maketitle

\begin{abstract}
  \noindent The {\em $k$-forest problem} is a common generalization of both the
  {\em $k$-MST} and the {\em dense-$k$-subgraph} problems. Formally,
  given a metric space on $n$ vertices $V$, with $m$ demand pairs
  $\subseteq V \times V$ and a ``target'' $k\le m$, the goal is to find
  a minimum cost subgraph that connects \emph{at least} $k$ demand
  pairs. In this paper, we give an
  $O(\min\{\sqrt{n},\sqrt{k}\})$-approximation algorithm for $k$-forest,
  improving on the previous best ratio of $O(n^{2/3}\log n)$ by Segev \&
  Segev~\cite{ss}.

  \medskip\noindent We then apply our algorithm for $k$-forest to obtain
   approximation algorithms for several {\em Dial-a-Ride} problems.
  The basic \dr~problem is the following: given an $n$ point metric space with
  $m$ objects each with its own source and destination, and a vehicle
  capable of carrying \emph{at most} $k$ objects at any time, find the
  minimum length tour that uses this vehicle to move each object from
  its source to destination. \ignore{ such that at most $k$ objects are
    in the vehicle at any time. We want that the tour be {\it
      non-preemptive}: i.e., each object, once picked up at its source,
    is dropped only at its destination.} We prove that an
  $\alpha$-approximation algorithm for the $k$-forest problem implies an
  $O(\alpha\cdot\log^2n)$-approximation algorithm for \dr. Using our
  results for $k$-forest, we get an
  $O(\min\{\sqrt{n},\sqrt{k}\}\cdot\log^2 n)$-approximation algorithm
  for \dr. The only previous result known for \dr was an $O(\sqrt{k}\log
  n)$-approximation by Charikar \& Raghavachari~\cite{cr}; our results
  give a different proof of a similar approximation guarantee---in fact,
  when the vehicle capacity $k$ is large, we give a slight improvement
  on their results. \ignore{The $\tilde{O}(\sqrt{k})$ part of our
    algorithm involves a reduction to the $k$-MST problem via an old
    theorem of Erd\H{o}s-Szekeres.}

  \medskip\noindent The reduction from \dr to the $k$-forest problem is fairly robust, and
  allows us to obtain approximation algorithms (with the same guarantee)
  for the following generalizations: (i) Non-uniform \dr, where the cost
  of traversing each edge is an arbitrary non-decreasing function of the
  number of objects in the vehicle; and (ii) Weighted \dr, where demands
  are allowed to have different weights. The reduction is essential,
  as it is unclear how to extend the techniques of Charikar \&
  Raghavachari to these \dr generalizations.

\end{abstract}

\ignore{
{\bf Keywords:} Approximation algorithms; Graph and network
algorithms; Vehicle routing
\end{titlepage}
\newpage}

\section{Introduction}

In the Steiner forest problem, we are given a set of vertex-pairs,
and the goal is to find a forest such that each vertex pair lies in
the same tree in the forest. This is a generalization of the Steiner
tree problem, where all the pairs contain a common vertex called the
root; both the tree and forest versions are well-understood
fundamental problems in network design~\cite{akr,gw}. An important
extension of the Steiner tree problem studied in the late 1990s was
the $k$-MST problem, where one sought the least-cost tree that
connected any $k$ of the terminals: several approximations
algorithms were given for the problem, culminating in the
$2$-approximation of Garg~\cite{garg}; the $k$-MST problem proved
crucial in many subsequent developments in network design and
vehicle routing~\cite{cgrt,fhr,bcklmm,bbcm}. One can analogously
define the $k$-forest problem where one needs to connect
\emph{only~$k$ of the pairs} in some Steiner forest instance:
surprisingly, very little is known about this problem, which was
first studied formally as recently as last year~\cite{hj,ss}. In
this paper, we give a simpler and improved approximation algorithm
for the \kfp.

Moreover, just like the $k$-MST variant, the \kfp seems to be useful
in applications to network design and vehicle routing. In the second
half of the paper, we show a (somewhat surprising) reduction of a
well-studied vehicle routing problem called the \dr problem to the
\kfp. In the \dr problem, we are given a metric space with people
having sources and destinations, and a bus of some capacity~$k$; the
goal is to find a route for this bus so that each person can be
taken from her source to destination without exceeding the capacity
of the bus at any point, such that the length of the bus route is
minimized. We show how the results for the \kfp slightly improve
upon existing results for the \dr problem; in fact, they give the
first approximation algorithms for some generalizations of \dr which
do not seem amenable to previous techniques.

\subsection{The $k$-Forest Problem}

Our starting point is the $k$-forest problem, which generalizes both the
$k$-MST and the dense-$k$-subgraph problems.
\begin{defn}[The $k$-Forest Problem]\label{kF-defn}
  Given an $n$-vertex metric space $(V,d)$, and \emph{demands} $\{s_i,
  t_i\}_{i=1}^m \sse V \times V$, find the least-cost subgraph that
  connects at least $k$ demand-pairs.
\end{defn}
Note that the \kfp is a generalization of the (minimization version
of the) well-studied \dks~problem, for which nothing better than an
$O(n^{1/3 - \delta})$ approximation is known.  The \kfp was first
defined in~\cite{hj}, and the first non-trivial approximation was
given by Segev and Segev~\cite{ss}, who gave an algorithm with an
approximation guarantee of $O(n^{2/3} \log n)$ for the case when $k
= O(\poly(n))$. We improve the approximation guarantee of the
$k$-forest problem to $O(\min\{\sqrt{n},\sqrt{k}\})$; formally, we
prove the following theorem in Section~\ref{kF-section}.
\begin{thm}[Approximating $k$-forest]
  \label{th:intro2}
  There is an $O(\min\{\sqrt{n}\cdot\frac{\log k}{\log n},
  \sqrt{k}\})$-approximation algorithm for the $k$-forest problem. For
  the case when $k$ is less than a polynomial in $n$, the approximation
  guarantee improves to $O(\min\{\sqrt{n},\sqrt{k}\})$.
\end{thm}

Apart from giving an improved approximation guarantee, our algorithm
for the \kfp is arguably simpler and more direct than that
of~\cite{ss} (which is based on Lagrangian relaxations for the
problem, and combining solutions to this relaxation). Indeed, we
give two algorithms, both reducing the \kfp to the \kmp in different
ways and achieving different approximation guarantees---we then
return the better of the two answers. The first algorithm (giving an
approximation of $O(\sqrt{k})$) uses the $k$-MST algorithm to find
good solutions on the sources and the sinks independently, and then
uses the Erd\H{o}s-Szekeres theorem on monotone subsequences to find
a ``good'' subset of these sources and sinks to connect cheaply;
details are given in Section~\ref{alg2}. The second algorithm starts
off with a single vertex as the initial solution, and uses the
$k$-MST algorithm to repeatedly find a low-cost tree that satisfies
a large number of demands which have one endpoint in the current
solution and the other endpoint outside; this tree is then used to
greedily augment the current solution and proceed. Choosing the
parameters (as described in Section~\ref{alg1}) gives us an
$O(\sqrt{n})$ approximation.

\subsection{The \dr Problem}
In this paper, we use the \kfp to give approximation algorithms for
the following vehicle routing problem.
\begin{defn}[The \dr Problem]\label{dr-defn}
  Given an $n$-vertex metric space $(V,d)$, a starting vertex (or
  \emph{root})~$r$, a set of $m$ demands $\{(s_i, t_i)\}_{i = 1}^m$, and a
  vehicle of capacity $k$, find a minimum length tour of the vehicle
  starting (and ending) at $r$ that moves each object $i$ from its
  source $s_i$ to its destination $t_i$ such that the vehicle carries at
  most $k$ objects at any point on the tour.
\end{defn}
We say that an object is {\it preempted} if, after being picked up from
its source, it can be left at some intermediate vertices before being
delivered to its destination. In this paper, we will not allow this, and
will mainly be concerned with the \emph{non-preemptive}
\dr~problem.\footnote{A note on the parameters: a feasible
  non-preemptive tour can be short-cut over vertices that do not
  participate in any demand, and we can assume that every vertex is an
  end point of some demand, and $n\le 2m$.  We may also assume, by
  preprocessing some demands, that $m\le n^2\cdot k$. However in
  general, the number of demands $m$ and the vehicle capacity $k$ may be
  much larger than the number of vertices $n$.}

The approximability of the \dr problem is not very well understood:
the previous best upper bound is an $O(\sqrt{k} \log
n)$-approximation algorithm due to Charikar and
Raghavachari~\cite{cr}, whereas the best lower bound that we are
aware of is APX-hardness (from TSP, say). We establish the following
(somewhat surprising) connection between the \dr and $k$-forest
problems in Section~\ref{dr-section}.
\begin{thm}[Reducing \dr to $k$-forest]
  \label{th:intro1}
  Given an $\alpha$-approximation algorithm for $k$-forest, there is an
  $O(\alpha\cdot\log^2 n)$-approximation algorithm for the \dr~problem.
\end{thm}
In particular, combining Theorems~\ref{th:intro2} and~\ref{th:intro1}
gives us an $O(\min\{\sqrt{k},\sqrt{n}\}\cdot \log^2n)$-approximation
guarantee for \dr. Of course, improving the approximation guarantee for
$k$-forest would improve the result for \dr as well.

Note that our results match the results of~\cite{cr} up to a
logarithmic term, and even give a slight improvement when the
vehicle capacity $k \gg n$, the number of nodes. Much more
interestingly, our algorithm for \dr easily extends to
generalizations of the \dr problem.  In particular, we consider a
substantially more general vehicle routing problem where the vehicle
has no \emph{a priori} capacity, and instead the cost of traversing
each edge $e$ is an arbitrary non-decreasing function $c_e(l)$ of
the number of objects $l$ in the vehicle; setting $c_e(l)$ to the
edge-length $d_e$ when $l \leq k$, and $c_e(l) = \infty$ for $l > k$
gives us back the classical \dr setting. In
Section~\ref{arbit-cost-section}, we show that this general {\em
non-uniform \dr} problem admits an approximation guarantee that
matches the best known for the classical \dr~problem. Another
extension we consider is the \emph{weighted \dr} problem. In this,
each object may have a different size, and total size of the items
in the vehicle must be bounded by the vehicle capacity; this has
been earlier studied as the {\it pickup and delivery}
problem~\cite{savs}. We show in Section~\ref{wt-section} that this
problem can be reduced to the (unweighted) \dr~problem at the loss
of only a constant factor in the approximation guarantee.

As an aside, we consider the effect of preemptions in the
\dr~problem (Section~\ref{pmt-section}). It was shown in Charikar \&
Raghavachari~\cite{cr} that the gap between the optimal preemptive
and non-preemptive tours could be as large as~$\Omega(n^{1/3})$. We
show that the real difference arises between \emph{zero} and
\emph{one} preemptions: allowing multiple preemptions does not give
us much added power. In particular, we show in Section~\ref{1-pmt}
that for any instance of the \dr~problem, there is a tour that
preempts each object \emph{at most once} and has length at most
$O(\log^2 n)$ times an optimal preemptive tour (which may preempt
each object an arbitrary number of times). Motivated by obtaining a
better guarantee for \dr on the Euclidean plane, we also study the
preemption gap in such instances. We show that even in this case,
there are instances having a gap of $\tilde{\Omega}(n^{1/8})$
between optimal preemptive and non-preemptive tours.

\subsection{Related Work}

\medskip\noindent{\bf The $k$-forest problem:} The $k$-forest problem is
relatively new: it was defined by Hajiaghayi \& Jain~\cite{hj}. An
$\tilde{O}(k^{2/3})$-approximation algorithm for even the directed
$k$-forest problem can be inferred from~\cite{ccc}.  Recently, Segev \&
Segev~\cite{ss} gave an $O(n^{2/3}\log n)$ approximation algorithm for
$k$-forest.

\medskip\noindent{\bf Dense $k$-subgraph:} The \emph{$k$-forest} problem
is a generalization of the \dks~problem~\cite{fpk}, as shown
in~\cite{hj}.  The best known approximation guarantee for the
\dks~problem is $O(n^{1/3-\delta})$ where $\delta>0$ is some constant,
due to Feige et al.~\cite{fpk}, and obtaining an improved guarantee has
been a long standing open problem.  Strictly speaking, Feige et
al.~\cite{fpk} study a potentially harder problem: the
\emph{maximization} version of \dks, where one wants to pick $k$
vertices to maximize the number of edges in the induced graph.  However,
nothing better is known even for the \emph{minimization} version of \dks
(where one wants to pick the minimum number of vertices that induce $k$
edges), which is a special case of $k$-forest. The $k$-forest problem is
also a generalization of $k$-MST, for which a 2-approximation is known
(Garg~\cite{garg}).

\medskip\noindent{\bf \dr:}
While the \dr problem has been studied extensively in the operations
research literature, relatively little is known about its
approximability. The currently best known approximation ratio for
\dr~is $O(\sqrt{k}\log n)$ due to Charikar \& Raghavachari
\cite{cr}. We note that their algorithm assumes instances with
unweighted demands. Krumke et al.~\cite{krw} give a 3-approximation
algorithm for the \dr~problem on a {\it line
  metric}; in fact, their algorithm finds a non-preemptive tour that has
length at most 3 times the preemptive lower bound. (Clearly, the cost of
an optimal preemptive tour is at most that of an optimal non-preemptive
tour.) A $2.5$-approximation algorithm for \emph{single source} version
of \dr (also called the ``capacitated vehicle routing'' problem) was
given by Haimovich \& Kan~\cite{hk}; again, their algorithm output a
non-preemptive tour with length at most 2.5 times the preemptive lower
bound. For the \emph{preemptive} \dr~problem, Charikar \&
Raghavachari~\cite{cr} gave the current-best $O(\log n)$ approximation
algorithm, and G{\o}rtz~\cite{g} showed that it is hard to approximate
this problem to better than $\Omega(\log^{1/4-\epsilon} n)$. Recall that
no super-constant hardness results are known for the non-preemptive \dr
problem.

\section{The $k$-forest problem}
\label{kF-section}

In this section, we study the $k$-forest problem, and give an
approximation guarantee of $O(\min\{\sqrt{n},\sqrt{k}\})$. This
result improves upon the previous best $O(n^{2/3}\log
n)$-approximation guarantee~\cite{ss} for this problem. The
algorithm in Segev \& Segev~\cite{ss} is based on a Lagrangian
relaxation for this problem, and suitably combining solutions to
this relaxation. In contrast, our algorithm uses a more direct
approach and is much simpler in description. Our approach is based
on approximating the following ``density'' variant of $k$-forest.
\begin{defn}[Minimum-ratio $k$-forest]
  \label{min-k-ratio}
  Given an $n$-vertex metric space $(V,d)$, $m$ pairs of vertices
  $\{s_i, t_i\}_{i=1}^m$, and a target $k$, find a tree $T$ that
  connects {\em at most} $k$ pairs, and minimizes the ratio of the
  length of $T$ to the number of pairs connected in $T$.\footnote{Even
    if we relax the solution to be any forest, we may assume (by
    averaging) that the \emph{optimal ratio} solution is a tree.}
\end{defn}
We present two different algorithms for \emph{minimum-ratio $k$-forest},
obtaining approximation guarantees of $O(\sqrt{k})$ (Section~\ref{alg2})
and $O(\sqrt{n})$ (Section~\ref{alg1}); these are then combined to give
the claimed result for the \kfp. Both our algorithms are based on subtle
reductions to the \kmp, albeit in very different ways.

As is usual, when we say that our algorithm {\em guesses} a parameter in
the following discussion, it means that the algorithm is run for each
possible value of that parameter, and the best solution found over all
the runs is returned. As long as only a constant number of parameters
are being guessed and the number of possibilities for each of these
parameters is polynomial, the algorithm is repeated only a polynomial
number of times.

\subsection{An $O(\sqrt{k})$ approximation algorithm}
\label{alg2}

In this section, we give an $O(\sqrt{k})$ approximation algorithm
for minimum ratio $k$-forest, which is based on a simple reduction
to the $k$-MST problem. The basic intuition is to look at the
solution $S$ to minimum-ratio $k$-forest and consider an Euler tour
of this tree $S$---a theorem of Erd\H{o}s \& Szekeres on increasing
subsequences implies that there must be at least $\sqrt{|S|}$
sources which are visited in the same order as the corresponding
sinks. We use this existence result to combine the source-sink pairs
to create an instance of $\sqrt{|S|}$-MST from which we can obtain a
good-ratio tree; the details follow.

Let $S$ denote an optimal ratio tree, that covers $q$ demands \& has
length $B$, and let $D$ denote the largest distance between any
demand pair that is covered in $S$ (note $D\le B$). We define a new
metric $l$ on the set $\{1,\cdots,m\}$ of demands as follows. The
distance between demands $i$ and $j$, $l_{i,j} =
d(s_i,s_j)+d(t_i,t_j)$, where $(V,d)$ is the original metric. The
$O(\sqrt{k})$ approximation algorithm first guesses the number of
demands $q$ \& the largest demand-pair distance $D$ in the optimal
tree $S$ (there are at most $m$ choices for each of $q$ \& $D$). The
algorithm discards all demand pairs $(s_i,t_i)$ such that
$d(s_i,t_i)>D$ (all the pairs covered in the optimal solution $S$
still remain). Then the algorithm runs the unrooted $k$-MST
algorithm~\cite{garg} with target $\lfloor\sqrt{q}\rfloor$, in the
metric $l$, to obtain a tree $T$ on the demand pairs $P$. From $T$,
we easily obtain trees $T_1$ (on all sources in $P$) and $T_2$ (on
all sinks in $P$) in metric $d$ such that $d(T_1)+d(T_2)=l(T)$.
Finally the algorithm outputs the tree $T'=T_1\cup T_2\cup \{e\}$,
where $e$ is any edge joining a source in $T_1$ to its corresponding
sink in $T_2$. Due to the pruning on demand pairs that have large
distance, $d(e)\le D$ and the length of $T'$, $d(T')\le l(T)+D\le
l(T)+B$.

We now argue that the cost of the solution $T$ found by the $k$-MST
algorithm $l(T)\le 8B$. Consider the optimal ratio tree $S$ (in metric
$d$) that has $q$ demands $\{(s_1,t_1),\cdots,(s_q,t_q)\}$, and let
$\tau$ denote an Euler tour of $S$. Suppose that in a traversal of
$\tau$, the {\it sources} of demands in $S$ are seen in the order
$s_1,\cdots,s_q$. Then in the same traversal, the {\it sinks} of demands
in $S$ will be seen in the order $t_{\pi(1)},\cdots,t_{\pi(q)}$, for
some permutation $\pi$. The following fact is well known (see,
e.g.,~\cite{steele-survey}).
\begin{thm}
  {\bf (Erd\H{o}s \& Szekeres)}\label{perm} Every permutation on
  $\{1,\cdots,q\}$ has either an increasing subsequence of length
  $\lfloor\sqrt{q}\rfloor$ or a decreasing subsequence of length
  $\lfloor\sqrt{q}\rfloor$.
\end{thm}
Using Theorem~\ref{perm}, we obtain a set $M$ of
$p=\lfloor\sqrt{q}\rfloor$ demands such that (1) the sources in $M$
appear in increasing order in a traversal of the Euler tour $\tau$,
and (2) the sinks in $M$ appear in increasing order in a traversal
of either $\tau$ or $\tau^R$ (the reverse traversal of $\tau$). Let
$j_0 < j_1<\cdots <j_{p-1}$ denote the demands in $M$ in increasing
order. From statement (1) above, $\sum_{i=0}^{p-1}
d(s(j_i),s(j_{i+1})) \le d(\tau)$, where the indices in the
summation are modulo $p$.  Similarly, statement (2) implies that
$\sum_{i=0}^{p-1} d(t(j_i),t(j_{i+1})) \le \max\{d(\tau),d(\tau^R)\}
=d(\tau)$. Thus we obtain:
$$\sum_{i=0}^{p-1} [d(s(j_i),s(j_{i+1})) + d(t(j_i),t(j_{i+1}))] \le
2d(\tau)\le 4B$$ But this sum is precisely the length of the tour
$j_0,j_1,\cdots,j_{p-1},j_0$ in metric $l$. In other words, there is
a tree of length $4B$ in metric $l$, that contains
$\lfloor\sqrt{q}\rfloor$ vertices. So, the cost of the solution $T$
found by the $k$-MST approximation algorithm is at most $8B$.

Now the final solution $T'$ has length at most $l(T)+B\le 9B$, and
ratio that at most $9\sqrt{q}\frac{B}{q}\le 9\sqrt{k}\frac{B}{q}$.
Thus we have an $O(\sqrt{k})$ approximation algorithm for minimum
ratio $k$-forest.

\subsection{An $O(\sqrt{n})$ approximation algorithm}
\label{alg1}

In this section, we show an $O(\sqrt{n})$ approximation algorithm
for the minimum ratio $k$-forest problem. The approach is again to
reduce to the $k$-MST problem; the intuition is rather different:
either we find a vertex $v$ such that a large number of demand-pairs
of the form $(v,*)$ can be satisfied using a small tree (the
``high-degree'' case); if no such vertex exists, we show that a
repeated greedy procedure would cover most vertices without paying
too much (and since we are in the ``low-degree'' case, covering most
vertices implies covering most demands too). The details follow.

Let $S$ denote an optimal solution to minimum ratio $k$-forest, and
$q\le k$ the number of demand pairs covered in $S$. We define the {\it
  degree} $\Delta$ of $S$ to be the maximum number of demands (among
those covered in $S$) that are incident at any vertex in $S$. The
algorithm first guesses the following parameters of the optimal solution
$S$: its length $B$ (within a factor 2), the number of pairs covered
$q$, the degree $\Delta$, and the vertex $w\in S$ that has $\Delta$
demands incident at it. Although, there may be an exponential number of
choices for the optimal length, a polynomial number of guesses within a
binary-search suffice to get a $B$ such that $B\le d(S)\le 2\cdot B$.
The algorithm then returns the better of the two procedures described
below.

\noindent {\bf Procedure 1 (high-degree case):} Since the degree of
vertex $w$ in the optimal solution $S$ is $\Delta$, there is tree
rooted at $w$ of length $d(S)\le 2B$, that contains at least
$\Delta$ demands having one end point at $w$. We assign a weight to
each vertex $u$, equal to the number of demands that have one end
point at this vertex $u$ and the other end point at $w$. Then we run
the $k$-MST algorithm~\cite{garg} with root $w$ and a target weight
of $\Delta$. By the preceding argument, this problem has a feasible
solution of length $2B$; so we obtain a solution $H$ of length at
most $4B$ (since the algorithm of \cite{garg} is a 2-approximation).
The ratio of solution $H$ is thus at most
$4B/\Delta=\frac{4q}{\Delta}\frac{B}{q}$.

\noindent {\bf Procedure 2 (low-degree case):} Set
$t=\frac{q}{2\Delta}$; note that $q\le \frac{\Delta\cdot n}{2}$ and
so $t\le n/4$. We maintain a current tree $T$ (initially just vertex
$w$), which is updated in iterations as follows: shrink $T$ to a
supernode $s$, and run the $k$-MST algorithm with root $s$ and a
target of $t$ new vertices. If the resulting $s$-tree has length at
most $4B$, include this tree in the current tree $T$ and continue.
If the resulting $s$-tree has length more than $4B$, or if all the
vertices have been included, the procedure ends. Since $t$ new
vertices are added in each iteration, the number of iterations is at
most $\frac{n}{t}$; so the length of $T$ is at most $\frac{4n}{t}B$.
We now show that $T$ contains at least $\frac{q}{2}$ demands.
Consider the set $S\setminus T$ (recall, $S$ is the optimal
solution). It is clear that $|S\setminus T|<t$; otherwise the
$k$-MST instance in the last iteration (with the current $T$) would
have $S$ as a feasible solution of length $\le 2B$ (and hence would
find one of length at most $4B$). So the number of demands covered
in $S$ that have at least one end point in $S\setminus T$ is at most
$|S\setminus T|\cdot \Delta\le t\cdot \Delta = q/2$ (as $\Delta$ is
the degree of solution $S$). Thus there are at least $q/2$ demands
{\em contained} in $S\cap T$, in particular in $T$. Thus $T$ is a
solution having ratio at most $\frac{4n}{t}B\cdot
\frac{2}{q}=\frac{8n}{t}\frac{B}{q}$.

The better ratio solution among $H$ and $T$ from the two procedures
has ratio at most $\min\{\frac{4q}{\Delta},
\frac{8n}{t}\}\cdot\frac{B}{q} =
\min\{8t,\frac{8n}{t}\}\cdot\frac{B}{q}\le
8\sqrt{n}\cdot\frac{B}{q}\le 8\sqrt{n}\cdot\frac{d(S)}{q}$. So this
algorithm is an $O(\sqrt{n})$ approximation to the minimum ratio
$k$-forest problem.

\subsection{Approximation algorithm for $k$-forest}
\label{ratio-to-kforest}

Given the two algorithms for minimum ratio $k$-forest, we can use
them in a standard greedy fashion (i.e., keep picking approximately
minimum-ratio solutions until we obtain a forest connecting at least
$k$ pairs); the standard set cover analysis can be used to show an
$O(\min\{\sqrt{n},\sqrt{k}\}\cdot\log k)$-approximation guarantee
for $k$-forest. A tighter analysis of the greedy algorithm (as done,
e.g., in Charikar et al.~\cite{ccc}) can be used to remove the
logarithmic terms and obtain the guarantee stated in
Theorem~\ref{th:intro2}.

\section{Applications to \dr~problems}
\label{dr-section}

In this section, we study applications of the $k$-forest problem to the
\dr problem (Definition~\ref{dr-defn}), and some generalizations. A
natural solution-structure for \dr involves servicing demands in batches
of at most $k$ each, where a batch consisting of a set $S$ of demands is
served as follows: the vehicle starts out being empty, picks up each of
the $|S|\le k$ objects from their sources, then drops off each object at
its destination, and is again empty at the end. If we knew that the
optimal solution has this structure, we could obtain a greedy framework
for \dr~by repeatedly finding the best `batch' of $k$ demands. However,
the optimal solution may involve carrying almost $k$ objects at every
point in the tour, in which case it can not be decomposed to be of the
above structure. In Theorem~\ref{pick-drop}, we show that there is
always a near optimal solution having this `pick-drop in batches'
structure. Building on Theorem~\ref{pick-drop}, we obtain approximation
algorithms for the classical \dr problem
(Section~\ref{basic-dr-section}), and two interesting extensions:
non-uniform \dr (Section~\ref{arbit-cost-section}) \& weighted \dr
(Section~\ref{wt-section}).

\begin{thm} [Structure Theorem]
\label{pick-drop} Given any instance of \dr, there exists a feasible
tour $\tau$ satisfying the following conditions:
\begin{enumerate}
 \item $\tau$ can be split into a
set of segments $\{S_1,\cdots ,S_t\}$ (i.e., $\tau=S_1\cdot
S_2\cdots S_t$) where each segment $S_i$ services a set $O_i$ of at
most $k$ demands such that $S_i$ is a path that first picks up each
demand in $O_i$ and then drops each of them.
 \item The length of $\tau$ is at most $O(\log m)$ times the length of
 an optimal tour.
\end{enumerate}
\end{thm}
\proof Consider an optimal non-preemptive tour $\sigma$: let
$c(\sigma)$ denote its length, and $|\sigma|$ denote the number of
edge traversals in $\sigma$. Note that if in some visit to a vertex
$v$ in $\sigma$ there is no pick-up or drop-off, then the tour can
be short-cut over vertex $v$, and it still remains feasible.
Further, due to triangle inequality, the length $c(\sigma)$ does not
increase by this operation. So we may assume that each vertex visit
in $\sigma$ involves a pick-up or drop-off of some object. Since
there is exactly one pick-up \& drop-off for each object, we have
$|\sigma|\le 2m+1$. Define the {\it stretch} of a demand $i$ to be
the number of edge traversals in $\sigma$ between the pick-up and
drop-off of object $i$. The demands are partitioned as follows: for
each $j=1,\cdots ,\lceil\log (2m)\rceil$, group $G_j$ consists of
all the demands whose stretch lie in the interval $[2^{j-1},2^j)$.
We consider each group $G_j$ separately.
\begin{cl}
\label{grp-tour} For each $j=1,\cdots ,\lceil\log (2m)\rceil$, there
is a tour $\tau_j$ that serves all the demands in group $G_j$,
satisfies condition~1 of Theorem~\ref{pick-drop}, and has length at
most $6\cdot c(\sigma)$.
\end{cl}
\proof Consider tour $\sigma$ as a line $\mathcal{L}$, with every
edge traversal in $\sigma$ represented by a distinct edge in
$\mathcal{L}$. Number the vertices in $\mathcal{L}$ from 0 to $h$,
where $h=|\sigma|$ is the number of edge traversals in $\sigma$.
Note that each vertex in $V$ may be represented multiple times in
$\mathcal{L}$. Each demand is associated with the numbers of the
vertices (in $\mathcal{L}$) where it is picked up \& dropped off.

Let $r=2^{j-1}$, and partition $G_j$ as follows: for $l=1,\cdots,
\lceil\frac{h}{r}\rceil$, set $O_{l,j}$ consists of all demands in
$G_j$ that are picked up at a vertex numbered between $(l-1)r$ and
$lr-1$. Since every demand in $G_j$ has stretch in the interval
$[r,2r]$, every demand in $O_{l,j}$ is dropped off at a vertex
numbered between $lr$ and $(l+2)r-1$. Note that $|O_{l,j}|$ equals
the number of demands in $G_j$ carried over edge $(lr-1,lr)$ by tour
$\sigma$, which is at most $k$. We define segment $S_{l,j}$ to start
at vertex number $(l-1)r$ and traverse all edges in $\mathcal{L}$
until vertex number $(l+2)r-1$ (servicing all demands in $O_{l,j}$
by first picking up each demand between vertices $(l-1)r$ \& $lr-1$;
then dropping off each demand between vertices $lr$ \& $(l+2)r-1$),
and then return (with the vehicle being empty) to vertex $lr$.
Clearly, the number of objects carried over any edge in $S_{l,j}$ is
at most the number carried over the corresponding edge traversal in
$\sigma$. Also, each edge in $\mathcal{L}$ participates in at most 3
segments $S_{l,j}$, and each edge is traversed at most twice in any
segment. So the total length of all segments $S_{l,j}$ is at most
$6\cdot c(\sigma)$. We define tour $\tau_j$ to be the concatenation
$S_{1,j}\cdots S_{\lceil h/r\rceil,j}$. It is clear that this tour
satisfies condition~1 of Theorem~\ref{pick-drop}.\bsq

Applying this claim to each group $G_j$, and concatenating the
resulting tours, we obtain the tour $\tau$ satisfying condition~1
and having length at most $6\log(2m)\cdot c(\sigma)=O(\log m)\cdot
c(\sigma)$.\bsq\\

\noindent {\bf Remark:} The ratio $O(\log m)$ in
Theorem~\ref{pick-drop} is almost best possible. There are instances
of \dr~(even on an unweighted line), where every solution satisfying
condition~1 of Theorem~\ref{pick-drop} has length at least
$\Omega(\max\{\frac{\log m}{\log\log m},\frac{k}{\log k}\})$ times
the optimal non-preemptive tour. So, if we only use solutions of
this structure, then it is not possible to obtain an approximation
factor (just in terms of capacity $k$) for \dr~that is better than
$\Omega(k/\log k)$. The solutions found by the algorithm for \dr~in
\cite{cr} also satisfy condition~1 of Theorem~\ref{pick-drop}. It is
interesting to note that when the underlying metric is a
hierarchically well-separated tree, \cite{cr} obtain a solution of
such structure having length $O(\sqrt{k})$ times the optimum,
whereas there is a lower bound of $\Omega(\frac{k}{\log k})$ even
for the simple case of an unweighted line.

\subsection{Classical Dial-a-Ride}
\label{basic-dr-section} Theorem~\ref{pick-drop} suggests a greedy
strategy for Dial-a-Ride, based on repeatedly finding the best batch
of $k$ demands to service. This greedy subproblem turns out to be
the minimum ratio $k$-forest problem (Definition~\ref{min-k-ratio}),
for which we already have an approximation algorithm. The next
theorem sets up the reduction from $k$-forest to \dr.

\begin{thm} [Reducing \dr~to minimum ratio $k$-forest] \label{dar-to-kforest}
A $\rho$-approximation algorithm for minimum ratio $k$-forest
implies an $O(\rho \log^2 m)$-approximation algorithm for
Dial-a-Ride.
\end{thm}
\proof The algorithm for  Dial-a-Ride is as follows.{\small
\begin{enumerate}
 \item $\mathcal{C}=\phi$.
 \item Until there are no uncovered demands, do:
 \begin{enumerate}
  \item Solve the minimum ratio $k$-forest problem, to obtain a
tree $C$ covering $k_C\le k$ new demands.
  \item Set $\mathcal{C}\leftarrow \mathcal{C}\cup C$.
 \end{enumerate}
\item For each tree $C\in\mathcal{C}$, obtain an Euler tour on $C$ to
locally service all demands (pick up all $k_C$ objects in the first
traversal, and drop them all in the second traversal). Then use a
1.5-approximate TSP tour on the sources, to connect all the local
tours, and obtain a feasible non-preemptive tour.
 \end{enumerate}}

Consider the tour $\tau$ and its segments as in
Theorem~\ref{pick-drop}. If the number of uncovered demands in some
iteration is $m'$, one of the segments in $\tau$ is a solution to
the minimum ratio $k$-forest problem of value at most
$\frac{d(\tau)}{m'}$. Since we have a $\rho$-approximation algorithm
for this problem, we would find a segment of ratio at most
$O(\rho)\cdot \frac{d(\tau)}{m'}$. Now a standard set cover type
argument shows that the total length of trees in $\mathcal{C}$ is at
most $O(\rho\log m)\cdot d(\tau)\le O(\rho\log^2 m)\cdot OPT$, where
$OPT$ is the optimal value of the Dial-a-Ride instance. Further, the
TSP tour on all sources is a lower bound on $OPT$, and we use a
1.5-approximate solution~\cite{c}. So the final non-preemptive tour
output in step~5 above has length at most $O(\rho\log^2 m)\cdot
OPT$.\bsq

\ignore{ It can be shown that a $\rho$-approximation for $k$-forest
implies a $\rho$-approximation for the greedy subproblem, and a
$\rho$-approximation for the greedy subproblem implies an
$O(\rho\cdot \log k)$-approximation for $k$-forest.

\begin{cor}
A $\rho$ approximation algorithm for $k$-forest implies an
$O(\rho\log^2 m)$ approximation algorithm for \dr.
\end{cor}
}

This theorem is in fact stronger than Theorem~\ref{th:intro1}
claimed earlier: it is easy to see that any approximation algorithm
for $k$-forest implies an algorithm with the same guarantee for
minimum ratio $k$-forest. Note that, $m$ and $k$ may be
super-polynomial in $n$. However, we show in
Section~\ref{wt-section} that with the loss of a constant factor,
the general \dr~problem can be reduced to one where the number of
demands $m\le n^4$. Based on this and Theorem~\ref{dar-to-kforest},
a $\rho$ approximation algorithm for minimum ratio $k$-forest
actually implies an $O(\rho\log^2 n)$ approximation algorithm for
\dr. Using the approximation algorithm for minimum ratio $k$-forest
(Section~\ref{kF-section}), we obtain an
$O(\min\{\sqrt{n},\sqrt{k}\}\cdot \log^2 n)$ approximation algorithm
for the \dr problem.

\noindent{\bf Remark:} If we use the $O(\sqrt{k})$ approximation for
$k$-forest, the resulting non-preemptive tour is in fact feasible
even for a $\sqrt{k}$ capacity vehicle! As noted in \cite{cr}, this
property is also true of their algorithm, which is based on an
entirely different approach.

\subsection{Non-uniform Dial-a-Ride} \label{arbit-cost-section}
The greedy framework for \dr described above is actually more
generally applicable than to just the classical \dr problem. In this
section, we consider the \dr~problem under a substantially more
general class of cost functions, and show how the $k$-forest problem
can be used to obtain an approximation algorithm for this
generalization as well. In fact, the approximation guarantee we
obtain by this approach matches the best known for the classical \dr
problem. Our framework for \dr is well suited for such a
generalization since it is a `primal' approach, based on directly
approximating a near-optimal solution; this approach is not too
sensitive to the cost function. On the other hand, the Charikar \&
Raghavachari~\cite{cr} algorithm is a `dual' approach, based on
obtaining a good lower bound, which depends heavily on the cost
function. Thus it is unclear whether their techniques can be
extended to handle such a generalization.

\begin{defn}[Non-uniform \dr]
Given an $n$ vertex undirected graph $G=(V,E)$, a root vertex $r$, a
set of $m$ demands $\{(s_i,t_i)\}_{i=1}^m$, and a non-decreasing
cost function $c_e:\{0,1,\cdots ,m\}\rightarrow \mathbb{R}^+$ on
each edge $e\in E$ (where $c_e(l)$ is the cost incurred by the
vehicle in traversing edge $e$ while carrying $l$ objects), find a
non-preemptive tour (starting \& ending at $r$) of minimum {\em
total cost} that moves each object $i$ from $s_i$ to $t_i$.
\end{defn}
Note that the classical \dr~problem is a special case when the edge
costs are given by: $c_e(l)=d_e$ if $l\le k$ \& $c_e(l)=\infty$
otherwise, where $d_e$ is the edge length in the underlying metric.
We may assume (without loss in generality) that for any fixed value
$l\in [0,m]$, the edge costs $c_e(l)$ induce a metric on $V$.
Similar to Theorem~\ref{pick-drop}, we have a near optimal solution
with a `batch' structure for the non-uniform \dr~problem as well,
which implies the algorithm in Theorem~\ref{non-unif-dr}. The proof
of the following corollary is almost identical to that of
Theorem~\ref{pick-drop}, and is omitted.

\begin{cor}[Non-uniform Structure Theorem] \label{pick-drop-g}
Given any instance of non-uniform \dr, there exists a feasible tour
$\tau$ satisfying the following conditions:
\begin{enumerate}
 \item $\tau$ can be split into a
set of segments $\{S_1,\cdots ,S_t\}$ (i.e., $\tau=S_1\cdot
S_2\cdots S_t$) where each segment $S_i$ services a set $O_i$ of
demands such that $S_i$ is a path that first picks up each demand in
$O_i$ and then drops each of them.
 \item The cost of $\tau$ is at most $O(\log m)$ times the cost of
 an optimal tour.
\end{enumerate}
\end{cor}

\begin{thm}[Approximating non-uniform \dr] \label{non-unif-dr}
A $\rho$-approximation algorithm for minimum ratio $k$-forest
implies an $O(\rho \log^2 m)$-approximation algorithm for
non-uniform \dr. In particular, there is an $O(\sqrt{n}\log^2
m)$-approximation algorithm.
\end{thm}
\proof Corollary~\ref{pick-drop-g} again suggests a greedy algorithm
for non-uniform \dr based on the following {\em greedy subproblem}:
find a set $T$ of uncovered demands and a path $\tau_0$ that first
picks up each object in $T$ and then drops off each of them, such
that the ratio of the cost of $\tau_0$ to $|T|$ is minimized.
However, unlike in the classical \dr~problem, in this case the cost
of path $\tau_0$ does not come from a single metric. Nevertheless,
the minimum ratio $k$-forest problem can be used to solve this
subproblem as follows.

{\small \begin{enumerate}
 \item For every $k=1,\cdots ,m$:
\begin{enumerate}
\item Define length function $d^{(k)}_e = c_e(k)$ on the edges.
\item Solve the minimum ratio $k$-forest problem on metric
$(V,d^{(k)})$ with target $k$, to obtain tree $T'_{k}$ covering
$n_k\le k$ demands.
\item Obtain an Euler tour $T_{k}$ of $T'_{k}$ that services these
$n_k$ demands, by picking up all demands in one traversal and then
dropping them all in a second traversal.
\end{enumerate}
 \item Return the tour $T_{k}$ having the smallest ratio
$\frac{c(T_k)}{n_k}$ (over all $1\le k\le m$). \end{enumerate}}

Assuming a $\rho$-approximation algorithm for minimum ratio
$k$-forest (for all values of $k$), we now show that the above
algorithm obtains a $16\rho$-approximate solution to the greedy
subproblem. The cost of tour $T_k$ in step~3 is $c(T_k)\le 4\cdot
d^{(k)}(T'_{k})$, since $T_k$ involves traversing a tour on tree
$T'_k$ twice and the vehicle carries at most $n_k\le k$ objects at
every point in $T_k$. So the ratio of tour $T_k$ is
$\frac{c(T_k)}{n_k}\le 4\frac{d^{(k)}(T'_{k})}{n_k}=4\cdot
\textrm{ratio}(T'_k)$. Let $\tau$ denote the optimal path for the
greedy subproblem,
 $T$ the set of demands that it services, and $t=|T|$. Let $T_1$ denote the
last $\frac{3}{4}t$ demands that are picked up, and $T_2$ denote the
first $\frac{3}{4}t$ demands that are dropped off. It is clear that
$T'=T_1\cap T_2$ has at least $t/2$ demands; let $T''\subset T'$ be
any subset with $|T''|=t/4$. Let $\tau'$ denote the portion of
$\tau$ between the $\frac{t}{4}$-th pick up and the
$\frac{3t}{4}$-th drop off. Note that when path $\tau$ is traversed,
there are at least $\frac{t}{4}$ objects in the vehicle while
traversing each edge in $\tau'$. So the cost of $\tau$, $c(\tau)\ge
\sum_{e\in \tau'} c_e(t/4)$. Since $\tau'$ contains the end points
of all demands in $T'\supset T''$, it is a feasible solution
(covering the demands $T''$) to minimum ratio $k$-forest with target
$k=t/4$ in the metric $d^{(t/4)}$, having ratio
$(\sum_{e\in\tau'}c_e(t/4))/\frac{t}{4}\le \frac{4c(\tau)}{t}$. So
the ratio of tour $T_{t/4}$ (obtained from the $\rho$-approximate
tree $T'_{t/4}$) is at most $4\cdot\textrm{ratio}(T'_k)\le
4\rho\frac{4c(\tau)}{t}=16\rho\frac{c(\tau)}{t}$. Thus we have a
$16\rho$-approximation algorithm for the greedy subproblem.

Based on Corollary~\ref{pick-drop-g}, it can now be shown (as in
Theorem~\ref{dar-to-kforest}) that a $\rho'$-approximation algorithm
for the greedy subproblem implies an $O(\rho'\cdot \log^2
m)$-approximation algorithm for non-uniform \dr. Using the above
$16\rho$-approximation for the greedy subproblem, we have the
theorem. \bsq

\ignore{\noindent {\bf Remark:} We observe that the preemptive
version of generalized \dr~can be solved using the recent algorithm
for the {\it non-uniform buy-at-bulk} problem~\cite{chks}. The
results of~\cite{chks} imply an $O(\log^4 m)$-approximation
algorithm for preemptive generalized \dr. The connection between
preemptive \dr~and the (uniform) buy-at-bulk problem was first shown
in G\o rtz~\cite{g}.}

\subsection{Weighted Dial-a-Ride} \label{wt-section} So
far we worked with the unweighted version of \dr, where each object
has the same weight. In this section, we extend our greedy framework
for \dr to the case when objects have different sizes, and the total
size of objects in the vehicle must be bounded by the vehicle
capacity. Here we only extend the classical \dr~problem and not the
generalization of Section~\ref{arbit-cost-section}. The problem
studied in this section has been studied earlier as the {\it pickup
and delivery} problem~\cite{savs}.
\begin{defn}[Weighted \dr]
Given a vehicle of capacity $Q\in\mathbb{N}$, an $n$-vertex metric
space $(V,d)$, a root vertex $r$, and a set of $m$ objects $\{(s_i,
t_i,w_i)\}_{i = 1}^m$ (with object $i$ having source $s_i$,
destination $t_i$ \& an integer size $1\le w_i\le Q$), find a
minimum length (non-preemptive) tour of the vehicle starting (and
ending) at $r$ that moves each object $i$ from its source to its
destination such that the total size of objects carried by the
vehicle is at most $Q$ at any point on the tour.
\end{defn}

The classical \dr~problem is a special case when $w_i=1$ for all
demands and the vehicle capacity $Q=k$. The following are two lower
bounds for weighted \dr: a TSP tour on the set of all sources \&
destinations (Steiner lower bound); and $\sum_{i=1}^{m}
\frac{w_i\cdot d(s_i,t_i)}{Q}$ (flow lower bound). In fact, as can
be seen easily, these two lower bounds are valid even for the
preemptive version of weighted \dr; so they are termed {\em
preemptive lower bounds}.

The main result of this section (Theorem~\ref{wt-algo}) reduces
weighted \dr to the classical \dr~problem with the additional
property that the number of demands ($m$) is small (polynomial in
the number of vertices $n$). This shows that in order to approximate
weighted \dr, it suffices to consider instances of the classical \dr
problem with a small number of demands. The next lemma shows that
even if the vehicle is allowed to split each object over multiple
deliveries, the resulting tour is {\em not} much shorter than the
tour where each object is required to be served in a single delivery
(as is the case in weighted \dr). This lemma is the main ingredient
in the proof of Theorem~\ref{wt-algo}. In the following, for any
instance of weighted \dr, we define the {\it unweighted instance}
corresponding to it as a classical \dr instance with vehicle
capacity $Q$, and $w_i$ (unweighted) demands each having source
$s_i$ and destination $t_i$ (for each $1\le i\le m$).
\begin{lem}
\label{unwt-to-wt} Given any instance $\mathcal{I}$ of weighted \dr,
and a solution $\tau$ to the unweighted instance corresponding to
$\mathcal{I}$, there is a polynomial time computable solution to
$\mathcal{I}$ having length at most $O(1)\cdot d(\tau)$.
\end{lem}
\proof Let $\mathcal{J}$ denote the unweighted instance
corresponding to $\mathcal{I}$. Define line $\mathscr{L}$ as in the
proof of Theorem~\ref{pick-drop} by traversing $\tau$ from $r$: for
every edge traversal in $\tau$, add a new edge of the same length at
the end of $\mathscr{L}$. For each unweighted object in
$\mathcal{J}$ corresponding to demand $i$ in $\mathcal{I}$, there is
a segment in $\tau$ (correspondingly in $\mathscr{L}$) where it is
moved from $s_i$ to $t_i$. So each demand $i\in\mathcal{I}$
corresponds to $w_i$ segments in $\tau$ (each being a path from
$s_i$ to $t_i$). For each demand $i$ in $\mathcal{I}$, we assign $i$
to one of its $w_i$ segments picked uniformly at random: call this
segment $l_i$. For an edge $e\in \mathscr{L}$, let $N_e=\sum_{i:e\in
l_i} w_i$ denote the random variable which equals the {\it total
weight} of demands whose assigned segments contain $e$. Note that
the expected value of $N_e$ is exactly the number of unweighted
objects carried by $\tau$ when traversing the edge corresponding to
$e$. Since $\tau$ is a feasible tour for $\mathcal{J}$, $E[N_e]\le
Q$ for all $e\in \mathscr{L}$.

Consider a random instance $\mathcal{R}$ of \dr~on line
$\mathscr{L}$ with vehicle capacity $Q$ and demands as follows: for
each demand $i$ in $\mathcal{I}$, an object of weight $w_i$ is to be
moved along segment $l_i$ (chosen randomly as above). Clearly, any
feasible tour for $\mathcal{R}$ corresponds to a feasible tour for
$\mathcal{I}$ of the same length. Note that the flow lower bound for
instance $\mathcal{R}$ is $F=\sum_{e\in \mathscr{L}} d_e\lceil
\frac{N_e}{Q}\rceil$, and the Steiner lower bound is $\sum_{e\in
\mathscr{L}} d_e = d(\tau)$. Using linearity of expectation,
$E[F]\le \sum_{e\in \mathscr{L}} d_e(\frac{E[N_e]}{Q}+1)\le 2\cdot
d(\tau)$. Let $R^*$ denote the instance  (on line $\mathscr{L}$)
obtained by assigning each demand $i$ in $\mathcal{I}$ to its
shortest length segment (among the $w_i$ segments corresponding to
it). Clearly this assignment minimizes the flow lower bound (over
all assignments of demands to segments). So $R^*$ has flow bound
$\le E[F]\le 2\cdot d(\tau)$, and Steiner lower bound $d(\tau)$.

Finally, we note that the 3-approximation algorithm for \dr~on a
line~\cite{krw} extends to a constant factor approximation algorithm
for the case with weighted demands as well (this can be seen
directly from~\cite{krw}). Additionally, this approximation
guarantee is relative to the preemptive lower bounds. Thus, using
this algorithm on $R^*$, we obtain a feasible solution to
$\mathcal{I}$ of length at most $O(1)\cdot d(\tau)$.\bsq

\begin{thm}[Weighted \dr~to unweighted]
\label{wt-algo} Suppose there is a $\rho$-approximation algorithm
for instances of classical \dr~with at most $O(n^4)$ demands. Then
there is an $O(\rho)$-approximation algorithm for weighted \dr (with
any number of demands). In particular, there is an $O(\sqrt{n}\log^2
n)$ approximation for weighted \dr.
\end{thm}
\proof Let $\mathcal{I}$ denote an instance of weighted \dr~with
objects $\{(w_i,s_i,t_i) : 1\le i\le m\}$, and $\tau^*$ an optimal
tour for $\mathcal{I}$. Let $\mathcal{P}=\{(s_1,t_1),\cdots
,(s_l,t_l)\}$ be the distinct pairs of vertices that have some
demand between them, and let $T_i$ denote the total size of {\em
all} objects having source $s_i$ and destination $t_i$. Note that
$l\le n(n-1)$. Let $\mathcal{P}_{high} = \{i\in\mathcal{P}:T_i\ge
\frac{Q}{2}\}$, $\mathcal{P}_{low} = \{i\in\mathcal{P}:T_i\le
\frac{Q}{l}\}$, and $\mathcal{P}'=\mathcal{P}\setminus
(\mathcal{P}_{high}\cup\mathcal{P}_{low})$. We now show how to
separately service objects in $\mathcal{P}_{low}$,
$\mathcal{P}_{high}$ \& $\mathcal{P}'$.

{\bf Servicing $\mathcal{P}_{low}$:} The total size in
$\mathcal{P}_{low}$ is at most $Q$; so we can service all these
pairs by traversing a single 1.5-approximate tour~\cite{c} on the
sources and destinations. Note that the length of this tour is at
most 1.5 times the Steiner lower bound, hence at most $1.5\cdot
d(\tau^*)$.

{\bf Servicing $\mathcal{P}_{high}$:} Let $C$ be a 1.5-approximate
minimum tour on all the sources. The pairs in $\mathcal{P}_{high}$
are serviced by a tour $\tau_1$ as follows. Traverse along $C$, and
when a source $s_i$ in $\mathcal{P}_{high}$ is visited, traverse the
direct edge to the corresponding destination $t_i$ \& back, as few
times as possible so as to move all the objects between $s_i$ and
$t_i$, as described next. Note that every object to be moved between
$s_i$ and $t_i$ has size (the original $w_i$ size) at most $Q$, and
the total size of such objects $T_i\ge Q/2$. So these objects can be
partitioned such that the size of each part (except possibly the
last) is in the interval $[\frac{Q}{2},Q]$. So the number of times
edge $(s_i,t_i)$ is traversed to service the demands between them is
at most $2\lceil\frac{2T_i}{Q}\rceil\le 2(\frac{2T_i}{Q}+1)\le
8\frac{T_i}{Q}$. Now, the length of tour $\tau_1$ is at most $d(C)+
\sum_{(s_i,t_i)\in \mathcal{P}_{high}} 8 d(s_i,t_i)\frac{T_i}{Q}\le
d(C)+ 8\sum_{i=1}^m \frac{w_i\cdot d(s_i,t_i)}{Q}$. Note that $d(C)$
is at most 1.5 times the minimum tour on all sources (Steiner lower
bound), and the second term above is the flow lower bound. So tour
$\tau_1$ has length at most $O(1)$ times the preemptive lower bounds
for $\mathcal{I}$, which is at most $O(1)\cdot d(\tau^*)$.

{\bf Servicing $\mathcal{P}'$:} We know that the total size $T_i$ of
each pair $i$ in $\mathcal{P}'$ lies in the interval $(Q/l,Q/2)$.
Let $\mathcal{I}'$ denote the instance of weighted \dr~with demands
$\{(s_i,t_i,T_i):i\in \mathcal{P}'\}$ and vehicle capacity $Q$; note
that the number of demands in $\mathcal{I}'$ is at most $l$. The
tour $\tau^*$ restricted to the objects corresponding to pairs in
$\mathcal{P}'$ is a feasible solution to the {\em unweighted
instance} corresponding to $\mathcal{I}'$ (but it may not feasible
for $\mathcal{I}'$ itself). However Lemma~\ref{unwt-to-wt} implies
that the optimal value of $\mathcal{I}'$, $opt(\mathcal{I})\le
O(1)\cdot d(\tau^*)$.

Next we reduce instance $\mathcal{I}'$ to an instance $\mathcal{J}$
of weighted \dr satisfying the following conditions: {\bf (i)}
$\mathcal{J}$ has at most $l$ demands, {\bf (ii)} each object in
$\mathcal{I}$ has size at most $2l$, {\bf (iii)} any feasible
solution to $\mathcal{J}$ is feasible for $\mathcal{I}'$, and {\bf
(iv)} the optimal value $opt(\mathcal{J})\le O(1)\cdot
opt(\mathcal{I}')$. If $Q\le 2l$, $\mathcal{J}=\mathcal{I}'$ itself
satisfies the required conditions. Suppose $Q\ge 2l$, then define
$p=\lfloor \frac{Q}{l}\rfloor$; note that $Q\ge l\cdot p\ge Q-l\ge
\frac{Q}{2}$. Round up each size $T_i$ to the smallest integral
multiple $T'_i$ of $p$, and round down the capacity $Q$ to
$Q'=l\cdot p$. Since each size $T_i\in (\frac{Q}{l},\frac{Q}{2})$,
all sizes $T'_i\in \{p,2p,\cdots, lp\}$. Now let $\mathcal{I}''$
denote the weighted \dr instance with demands $\{(s_i,t_i,T'_i):i\in
\mathcal{P}'\}$ and vehicle capacity $Q'=lp$. One can obtain a
feasible solution for $\mathcal{I}''$ from any feasible solution
$\sigma$ for $\mathcal{I}'$ by traversing $\sigma$ a constant number
of times: this follows from $Q'\ge \frac{Q}{2}$ \& $T'_i\le \max\{
2T_i,Q'\}$.\footnote{In particular, consider simulating a traversal
along $\sigma$ of a capacity $Q$ vehicle ($T_0$) by 8 capacity $Q'$
vehicles $T'_1,\cdots,T'_8$, each running in parallel along
$\sigma$. Whenever vehicle $T_0$ picks-up an object $i$, one of the
vehicles $\{T'_g\}_{g=1}^8$ picks-up $i$: if $w_i\le \frac{Q}{4}$,
any vehicle $\{T'_g\}_{g=1}^4$ that has free capacity picks-up $i$;
if $w_i>\frac{Q}{4}$, any vehicle $\{T'_g\}_{g=5}^8$ that is empty
picks-up $i$. It is easy to see that if at some point none of the
vehicles $\{T'_g\}_{g=1}^8$ picks-up an object, there must be a
capacity violation in $T_0$.}~ So the optimal value of
$\mathcal{I}''$ is at most $O(1)\cdot opt(\mathcal{I}')$. Now note
that all sizes and the vehicle capacity in $\mathcal{I}''$ are
multiples of $p$; scaling down each of these quantities by $p$, we
get an instance $\mathcal{J}$ {\it equivalent} to $\mathcal{I}''$
where the vehicle capacity is $l$ (and every demand size is at most
$l$). This instance $\mathcal{J}$ satisfies all the four conditions
claimed above.

Now observe that the instance $\mathcal{J}$ can be solved using
$\rho$-approximation algorithm assumed in the theorem. Since
$\mathcal{J}$ has at most $l$ demands (each of size $\le 2l$), the
unweighted instance corresponding to $\mathcal{J}$ has at most
$2l^2\le 2n^4$ demands. Thus, this unweighted instance can be solved
using the $\rho$-approximation algorithm for such instances, assumed
in the theorem. Then using the algorithm in Lemma~\ref{unwt-to-wt},
we obtain a solution to $\mathcal{J}$, of length at most
$O(\rho)\cdot opt(\mathcal{J})\le O(\rho)\cdot opt(\mathcal{I}')\le
O(\rho)\cdot d(\tau^*)$. Since any feasible solution to
$\mathcal{J}$ corresponds to one for $\mathcal{I}'$, we have a tour
servicing $\mathcal{P}'$ of length at most $O(\rho)\cdot d(\tau^*)$.

Finally, combining the tours servicing $\mathcal{P}_{low}$,
$\mathcal{P}_{high}$ \& $\mathcal{P}'$, we obtain a feasible tour
for $\mathcal{I}$ having length $O(\rho)\cdot d(\tau^*)$, which
gives us the desired approximation algorithm. \bsq
\\

\noindent Theorem~\ref{wt-algo} also justifies the assumption $\log
m=O(\log n)$ made at the end of Section~\ref{dr-section}. This is
important because in general $m$ may be super-polynomial in $n$.

\section{The Effect of Preemptions}
\label{pmt-section} In this section, we study the effect of the
number of preemptions in the \dr~problem. We mentioned two versions
of the \dr~problem (Definition~\ref{dr-defn}): in the preemptive
version, an object may be preempted any number of times, and in the
non-preemptive version objects are not allowed to be preempted even
once. Clearly the preemptive version is least restrictive and the
non-preemptive version is most restrictive. One may consider other
versions of the \dr~problem, where there is a specified upper bound
$P$ on the number of times an object can be preempted. Note that the
case $P=0$ is the non-preemptive version, and the case $P=n$ is the
preemptive version. We show that for any instance of the
\dr~problem, there is a tour that preempts each object at most once
(i.e., $P=1$) and has length at most $O(\log^2 n)$ times an optimal
preemptive tour (i.e., $P=n$). This implies that the real gap
between preemptive and non-preemptive tours is between zero and one
preemption per object. A tour that preempts each object at most once
is called a {\it 1-preemptive tour}.
\begin{thm}[Many preemptions to one preemption]
\label{1-pmt} Given any instance of the \dr problem, there is a
1-preemptive tour of length at most $O(\log^2 n)\cdot OPT_{pmt}$,
where $OPT_{pmt}$ is the length of an optimal preemptive tour. Such
a tour can be found in randomized polynomial time.
\end{thm}
\proof Using the results on probabilistic tree embedding~\cite{frt},
we may assume that the given metric is a {\em hierarchically
well-separated} tree $T$. This only increases the expected length of
the optimal solution by a factor of $O(\log n)$. Further, tree $T$
has $O(\log \frac{d_{max}}{d_{min}})$ levels, where $d_{max}$ and
$d_{min}$ denote the maximum and minimum distances in the original
metric. We first observe that using standard scaling arguments, it
suffices to assume that $\frac{d_{max}}{d_{min}}$ is polynomial in
$n$. Without loss of generality, any preemptive tour involves at
most $2m\cdot n$ edge traversals: each object is picked or dropped
at most $2n$ times (once at each vertex), and every visit to a
vertex involves picking or dropping at least one object (otherwise
the tour can be shortcut over this vertex at no increase in length).
By retaining only vertices within distance $OPT_{pmt}/2$ from the
root $r$, we preserve the optimal preemptive tour and ensure that
$d_{max}\le OPT_{pmt}$. Now consider modifying the original metric
by setting all edges of length smaller than $OPT_{pmt}/2mn^3$ to
length 0; the new distances are shortest paths under the modified
edge lengths. So any pairwise distance decreases by at most
$\frac{OPT_{pmt}}{2mn^2}$. Clearly the length of the optimal
preemptive tour only decreases under this modification. Since there
are at most $2mn$ edge traversals in any preemptive tour, the
increase in tour length in going from the new metric to the original
metric is at most $2mn\cdot \frac{OPT_{pmt}}{2mn^2}\le
\frac{OPT_{pmt}}{n}$. Thus at the loss of a constant factor, we may
assume that $d_{max}/d_{min}\le 2mn^3$. Further, the reduction in
Theorem~\ref{unwt-to-wt} also holds for preemptive \dr; so we may
assume (at the loss of an additional constant factor) that the
number of demands $m\le O(n^4)$. So we have $d_{max}/d_{min}\le
O(n^7)$ and hence tree $T$ has $O(\log n)$ levels.

The tree $T$ resulting from the probabilistic embedding has several
Steiner vertices that are not present in the original metric; so the
tour that we find on $T$ may actually preempt objects at Steiner
vertices, in which case it is not feasible in the original metric.
However as shown by Gupta~\cite{gupta}, these Steiner vertices can
be simulated by vertices in the original metric (at the loss of a
constant factor). Based on the preceding observations, we assume
that the metric is a tree $T$ on the original vertex set having
$l=O(\log n)$ levels, such that the expected length of the optimal
preemptive tour is $O(\log n)\cdot OPT_{pmt}$.

We now partition the demands in $T$ into $l$ sets with $D_i$ (for
$i=1,\cdots,l$) consisting of all demands having their least common
ancestor (lca) in level $i$. We service each $D_i$ separately using
a tour of length $O(OPT_{pmt})$. Then concatenating the tours for
each level $i$, we obtain the theorem.

\noindent {\bf Servicing $D_i$:} For each vertex $v$ at level $i$ in
$T$, let $L_v$ denote the demands in $D_i$ that have $v$ as their
lca. Consider an optimal {\it preemptive} tour that services the
demands $D_i$. Since the subtrees under any two different level $i$
vertices are disjoint and there is no demand in $D_i$ across such
subtrees, we may assume that this optimal tour is a concatenation of
disjoint preemptive tours servicing each $L_v$ separately. If
$OPT_{pmt}(v)$ denotes the length of an optimal preemptive tour
servicing $L_v$ with $v$ as the starting vertex, $\sum_v
OPT_{pmt}(v) \le OPT_{pmt}$.

Now consider an optimal preemptive tour $\tau_v$ servicing $L_v$.
Since the $s_j-t_j$ path of each demand $j\in L_v$ crosses vertex
$v$, at some point in tour $\tau_v$ the vehicle is at $v$ with
object $j$ in it. Consider the tour $\sigma_v$ obtained by modifying
$\tau_v$ so that it drops each object $j$ at $v$ when the vehicle is
at $v$ with object $j$ in it. Clearly
$d(\sigma_v)=d(\tau_v)=OPT_{pmt}(v)$. Note that $\sigma_v$ is a
feasible preemptive tour for the {\em single source \dr}~problem
with sink $v$ and all sources in $L_v$. Thus the algorithm of
\cite{hk} gives a non-preemptive tour $\sigma'_v$ that moves all
objects in $L_v$ from their sources to $v$, having length at most
$2.5d(\sigma_v)=2.5OPT_{pmt}(v)$. Similarly, we can obtain a
non-preemptive tour $\sigma''_v$ that moves all objects in $L_v$
from $v$ to their destinations, having length at most
$2.5OPT_{pmt}(v)$. Now $\sigma'_v\cdot \sigma''_v$ is a 1-preemptive
tour servicing $L_v$ of length at most $5\cdot OPT_{pmt}(v)$.

We now run a DFS on $T$ to visit all vertices in level $i$, and use
the algorithm described above for servicing demands $L_v$ when $v$
is visited in the DFS. This results in a tour servicing $D_i$,
having length at most $2d(T) + 5\sum_v OPT_{pmt}(v)$. Here $2d(T)$
is the Steiner lower bound, and $\sum_v OPT_{pmt}(v) \le OPT_{pmt}$.
Thus the tour servicing $D_i$ has length at most $6\cdot OPT_{pmt}$.

Finally concatenating the tours for each level $i=1,\cdots,l$, we
obtain a 1-preemptive tour on $T$ of length $O(\log n)\cdot
OPT_{pmt}$, which translates to a 1-preemptive tour on the original
metric having length $O(\log^2 n)\cdot OPT_{pmt}$. \bsq

Motivated by obtaining an improved approximation for \dr on the
Euclidean plane, we next consider the worst case gap between an
optimal non-preemptive tour and the preemptive lower bounds. As
mentioned earlier,~\cite{cr} showed that there are instances of
\dr~where the ratio of the optimal non-preemptive tour to the
optimal preemptive tour is $\Omega(n^{1/3})$. However, the metric
involved in this example was the uniform metric on $n$ points, which
can not be embedded in the Euclidean plane. The following theorem
shows that even in this special case, there can be a polynomial gap
between non-preemptive and preemptive tours, and implies that just
preemptive lower bounds do not suffice to obtain a poly-logarithmic
approximation guarantee.
\begin{thm}[Preemption gap in Euclidean plane]
\label{plane-lb} There are instances of \dr~on the Euclidean plane
where the optimal non-preemptive tour has length
$\Omega(\frac{n^{1/8}}{\log^3 n})$ times the optimal preemptive
tour.
\end{thm}
\proof Consider a square of side 1 in the Euclidean plane, in which
a set of $n$ demand pairs are distributed uniformly at random (each
demand point is generated independently and is distributed uniformly
at random in the square). The vehicle capacity is set to
$k=\sqrt{n}$. Let $\mathcal{R}$ denote a random instance of
\dr~obtained as above. We show that in this case, the optimal
non-preemptive tour has length $\tilde{\Omega}(n^{1/8})$ with high
probability. We first show the following claim.
\begin{cl}
\label{rand-k-tree} The minimum length of a tree containing $k$
pairs in $\mathcal{R}$ is $\Omega(\frac{n^{1/8}}{\log n})$, w.h.p.
\end{cl}
\proof Take any set $S$ of $k=\sqrt{n}$ demand pairs. Note that the
number of such sets $S$ is ${n \choose k}$. This set $S$ has $2k$
points each of them generated uniformly at random. It is known that
there are $p^{p-2}$ different labeled trees on $p$ vertices (see
e.g.~\cite{vw}, Ch.2). The term {\em labeled} emphasizes that we are
not  identifying isomorphic graphs, i.e., two trees are counted as
the  same if and only if exactly the same pairs of vertices are
adjacent. Thus there are at most $(2k)^{2k- 2}$ such trees just on
set $S$. Consider any tree $T$ among these trees and root it at the
source point with minimum label. Here we assume  that $T$ has been
generated using the ``Principle of Deferred Decisions'', i.e., nodes
will be generated one by one according to some breadth-first
ordering of $T$. We say that an edge is {\it short} if its length is
at most $\frac{c}{\alpha k}$ ($c$ and $\alpha\in (0,\frac{1}{2})$
will be fixed later).

If $T$ has length at most $c$, it is clear that at most an $\alpha$
fraction of its edges are {\it not} short. So $Pr[length(T)\le c]
\le \sum_{H} Pr[edges~in~H~are~short]$, where $H$ in the summation
ranges over all edge-subsets in $T$ with $|H|\ge (1-\alpha)2k$. For
a fixed $H$, we bound $Pr[edges~in~H~are~short]$ as follows. For any
edge $(v,\text{parent}(v))$ (note $\text{parent}(v)$ is well-defined
since $T$ is  rooted), assuming that $\text{parent}(v)$ is fixed,
the probability that this edge is short is $p=\pi(\frac{c}{\alpha
k})^2$. So we can upper bound the probability that edges $H$ are
short by $p^{|H|}\le p^{(1-\alpha)2k}$. So we have $Pr[length(T)\le
c] \le 2^{2k}\cdot p^{(1-\alpha)2k}$, as the number of different
edge sets $H$ is at most $2^{2k}$.

By a union bound over all such labeled trees $T$, the probability
that the length of the minimum spanning tree on $S$ is less than $c$
is at most $(2k)^{2k}\cdot2^{2k}\cdot p^{(1-\alpha)2k}$. Now taking
a union bound over all $k$-sets $S$, the probability that the
minimum length of a tree containing $k$ pairs is less than $c$ is at
most ${n \choose k}(2k)^{2k}2^{2k}p^{(1-\alpha)2k}$. Since
$k=\sqrt{n}$, this term can be bounded as follows:
$$
(ek)^{k} (4k)^{2k} \pi^{(1-\alpha)2k} (\frac{c}{\alpha
k})^{{(1-\alpha)4k}} \le  500^k k^{3k} (\frac{c}{\alpha
k})^{(1-\alpha)4k} =[500\cdot (\frac{c}{\alpha})^{4-4\alpha}
(\frac{1}{k})^{1-4\alpha}]^k \le 2^{-k}
$$
The last inequality above holds when $c\le \frac{\alpha}{1000}\cdot
k^{1/4-3\alpha/(1-4\alpha)}$. Setting $\alpha = \frac{1}{\log k}$,
we get $$Pr[\exists~~\frac{k^{1/4}}{8000\cdot\log k} \textrm{ length
tree containing $k$ pairs in $\mathcal{R}$}] \le 2^{-k}$$ So, with
probability at least $1-2^{-\sqrt{n}}$, the minimum length of a tree
containing $k$ pairs in $\mathcal{R}$ is at least
$\Omega(\frac{n^{1/8}}{\log n})$. \bsq

From Theorem~\ref{pick-drop}, we obtain that there is a near optimal
non-preemptive tour servicing all the demands in segments, where
each segment (except possibly the last) involves servicing a set of
$\frac{k}{2}\le t \le k$ demands. Although the lower bound of $k/2$
is not stated in Theorem~\ref{pick-drop}, it is easy to extend the
statement to include it. This implies that any solution of this
structure has at least $\frac{n}{k}=k$ segments. Since each segment
covers at least $k/2$ pairs, Claim~\ref{rand-k-tree} implies that
each of these segments has length $\Omega(n^{1/8}/\log n)$. So the
best solution of the structure given in Theorem~\ref{pick-drop} has
length $\Omega(\frac{n^{1/8}}{\log n}k)$. But since there is a
near-optimal solution of this structure, the optimal non-preemptive
tour on $\mathcal{R}$ has length $\Omega(\frac{n^{1/8}}{\log^2
n}k)$.

On the other hand, the flow lower bound for $\mathcal{R}$ is at most
$\frac{n}{k}=k$, and the Steiner lower bound is at most
$O(\sqrt{n})=O(k)$ (an $O(\sqrt{n})$ length tree on the $2n$ points
can be constructed using a $\sqrt{2n}\times \sqrt{2n}$ gridding). So
the preemptive lower bounds are both $O(k)$; now using the algorithm
of~\cite{cr}, we see that the optimal preemptive tour has length
$O(k\log n)$. Combined with the lower bound for non-preemptive
tours, we obtain the Theorem.\bsq

\ignore{\section{Conclusions} It would be very interesting to obtain
a poly-logarithmic approximation for \dr, even on the Euclidean
plane. One way to do this could be to improve the $k$-forest
algorithm in this setting (this restriction does not contain the
dense-$k$-subgraph problem). However even in the Euclidean plane, we
have shown that there are instances of \dr~where the gap between the
optimal non-preemptive and preemptive tours is
$\tilde{\Omega}(n^{1/8})$. So this would require using better lower
bounds than just the preemptive ones.

Our algorithm for \dr~basically reduces the problem to $k$-forest.
It would be interesting to determine if the reverse reduction holds.
This would show that the \dr~problem is as hard to approximate as
the dense-$k$-subgraph problem, which is believed to be hard to
approximate better than a polynomial factor~\cite{st}.}

\bigskip
\noindent{\bf Acknowledgements:} We thank Alan Frieze for his help
in proving Theorem~\ref{plane-lb}.

\bibliography{dial-ride}

\newcommand{\etalchar}[1]{$^{#1}$}
\begin{thebibliography}{CCyC{\etalchar{+}}98}

\bibitem[AKR91]{akr}
Ajit Agrawal, Philip Klein, and R.~Ravi.
\newblock When trees collide: an approximation algorithm for the generalized
  steiner problem on networks.
\newblock {\em Proceedings of the 23rd Annual ACM Symposium on Theory of
  Computing}, pages 134--144, 1991.

\bibitem[BBCM04]{bbcm}
N.~Bansal, A.~Blum, S.~Chawla, and A.~Meyerson.
\newblock {Approximation Algorithms for Deadline-TSP and Vehicle Routing with
  Time Windows}.
\newblock {\em Proceedings of the 36th Annual ACM Symposium on Theory of
  Computing}, pages 166--174, 2004.

\bibitem[BCK{\etalchar{+}}03]{bcklmm}
A.~Blum, S.~Chawla, D.~R. Karger, T.~Lane, A.~Meyerson, and M.~Minkoff.
\newblock {Approximation Algorithms for Orienteering and Discounted-Reward
  TSP}.
\newblock {\em Proceedings of the 44th Annual IEEE Symposium on Foundations of
  Computer Science}, pages 46--55, 2003.

\bibitem[CCyC{\etalchar{+}}98]{ccc}
Moses Charikar, Chandra Chekuri, To~yat Cheung, Zuo Dai, Ashish Goel, Sudipto
  Guha, and Ming Li.
\newblock {Approximation algorithms for directed Steiner problems}.
\newblock In {\em SODA '98: Proceedings of the ninth annual ACM-SIAM symposium
  on Discrete algorithms}, pages 192--200, 1998.

\bibitem[CGRT03]{cgrt}
Kamalika Chaudhuri, Brighten Godfrey, Satish Rao, and Kunal Talwar.
\newblock Paths, trees, and minimum latency tours.
\newblock {\em Proceedings of the 44th Annual IEEE Symposium on Foundations of
  Computer Science}, pages 36--45, 2003.

\bibitem[Chr77]{c}
N.~Christofides.
\newblock {Worst-case analysis of a new heuristic for the travelling salesman
  problem}.
\newblock {\em GSIA, CMU-Report 388}, 1977.

\bibitem[CR98]{cr}
Moses Charikar and Balaji Raghavachari.
\newblock {The Finite Capacity Dial-A-Ride Problem}.
\newblock In {\em {IEEE} Symposium on Foundations of Computer Science}, pages
  458--467, 1998.

\bibitem[FHR03]{fhr}
Jittat Fakcharoenphol, Chris Harrelson, and Satish Rao.
\newblock The k-traveling repairman problem.
\newblock {\em Proceedings of the 14th annual ACM-SIAM symposium on Discrete
  algorithms}, pages 655--664, 2003.

\bibitem[FPK01]{fpk}
Uriel Feige, David Peleg, and Guy Kortsarz.
\newblock {The Dense k -Subgraph Problem}.
\newblock {\em Algorithmica}, 29(3):410--421, 2001.

\bibitem[FRT03]{frt}
Jittat Fakcharoenphol, Satish Rao, and Kunal Talwar.
\newblock {A tight bound on approximating arbitrary metrics by tree metrics}.
\newblock In {\em STOC '03: Proceedings of the thirty-fifth annual ACM
  symposium on Theory of computing}, pages 448--455, 2003.

\bibitem[Gar05]{garg}
Naveen Garg.
\newblock {Saving an epsilon: a 2-approximation for the k-MST problem in
  graphs}.
\newblock In {\em Proceedings of the thirty-seventh annual ACM symposium on
  Theory of computing}, pages 396--402, 2005.

\bibitem[Gup01]{gupta}
Anupam Gupta.
\newblock Steiner points in tree metrics don't (really) help.
\newblock In {\em SODA '01: Proceedings of the twelfth annual ACM-SIAM
  symposium on Discrete algorithms}, pages 220--227, 2001.

\bibitem[GW92]{gw}
Michel~X. Goemans and David~P. Williamson.
\newblock A general approximation technique for constrained forest problems.
\newblock {\em Proceedings of the 3rd Annual ACM-SIAM Symposium on Discrete
  Algorithms}, pages 307--316, 1992.

\bibitem[HJ06]{hj}
Mohammad~Taghi Hajiaghayi and Kamal Jain.
\newblock {The prize-collecting generalized steiner tree problem via a new
  approach of primal-dual schema}.
\newblock In {\em SODA '06: Proceedings of the seventeenth annual ACM-SIAM
  symposium on Discrete algorithm}, pages 631--640, 2006.

\bibitem[HK85]{hk}
M.~Haimovich and A.~H. G.~Rinnooy Kan.
\newblock {Bounds and heuristics for capacitated routing problems}.
\newblock {\em Mathematics of Operations Research}, 10:527--542, 1985.

\bibitem[KRW00]{krw}
S.~Krumke, J.~Rambau, and S.~Weider.
\newblock {An approximation algorithm for the nonpreemptive capacitated
  dial-a-ride problem}.
\newblock {\em Preprint 00-53, Konrad-Zuse-Zentrum fr Informationstechnik
  Berlin}, 2000.

\bibitem[rtz06]{g}
Inge Li~G\o rtz.
\newblock {Hardness of Preemptive Finite Capacity Dial-a-Ride}.
\newblock In {\em 9th International Workshop on Approximation Algorithms for
  Combinatorial Optimization Problems}, 2006.

\bibitem[SS95]{savs}
M.W.P. Savelsbergh and M.~Sol.
\newblock {The general pickup and delivery problem}.
\newblock {\em Transportation Science}, 29:17--29, 1995.

\bibitem[SS06]{ss}
Danny Segev and Gil Segev.
\newblock {Approximate k-Steiner Forests via the Lagrangian Relaxation
  Technique with Internal Preprocessing}.
\newblock In {\em 14th Annual European Symposium on Algorithms}, pages
  600--611, 2006.

\bibitem[Ste95]{steele-survey}
J.~Michael Steele.
\newblock Variations on the monotone subsequence theme of {E}rd{\H o}s and
  {S}zekeres.
\newblock In {\em Discrete probability and algorithms (Minneapolis, MN, 1993)},
  volume~72 of {\em IMA Vol. Math. Appl.}, pages 111--131. Springer, New York,
  1995.

\bibitem[vLW92]{vw}
J.~H. van Lint and R.~M. Wilson.
\newblock {\em {A Course in Combinatorics}}.
\newblock Cambridge University Press, 1992.

\end{thebibliography}
\bibliographystyle{alpha}

\end{document}